# Mapping Dirac Quasiparticles near a Single Coulomb Impurity on Graphene

Yang Wang[1,2*], Victor W. Brar[1,2*], Andrey V. Shytov[3], Qiong Wu[1,2], William Regan[1,2], Hsin-Zon Tsai[1], Alex Zettl[1,2], Leonid S. Levitov[4] and Michael F. Crommie[1,2]

[1]*Department of Physics, University of California at Berkeley, Berkeley CA, 94720, United States.*

[2]*Materials Science Division, Lawrence Berkeley National Laboratory, Berkeley CA, 94720, United States.*

[3]*School of Physics, University of Exeter, Stocker Road, Exeter EX4 4QL, United Kingdom.*

[4]*Department of Physics, Massachusetts Institute of Technology, 77 Massachusetts Ave, Cambridge MA, 02139, United States.*

*These authors contributed equally to this work

**The response of Dirac fermions to a Coulomb potential is predicted to differ significantly from the behavior of non-relativistic electrons seen in traditional atomic and impurity systems[1-3]. Surprisingly, many key theoretical predictions for this ultra-relativistic regime have yet to be tested in a laboratory[4-12]. Graphene, a 2D material in which electrons behave like massless Dirac fermions[13-15], provides a unique opportunity to experimentally test such predictions. The response of Dirac fermions to a Coulomb potential in graphene is central to a wide range of electronic phenomena and can serve as a sensitive probe of graphene's intrinsic dielectric constant [6,8], the primary factor determining the strength of electron-electron interactions in this material[16]. Here we present a direct measurement of the nanoscale response of Dirac fermions to a single Coulomb potential placed on a gated graphene device. Scanning tunneling microscopy and spectroscopy were used to fabricate tunable charge impurities on graphene and to measure how they are screened by Dirac fermions for a $Q = +1|e|$ impurity charge state. Electron-like and hole-like Dirac fermions were observed to respond very differently to tunable Coulomb potentials. Comparison of this electron-hole asymmetry to theoretical**

**simulations has allowed us to test basic predictions for the behavior of Dirac fermions near a Coulomb potential and to extract the intrinsic dielectric constant of graphene: $\varepsilon_g = 3.0 \pm 1.0$. This small value of $\varepsilon_g$ indicates that microscopic electron-electron interactions can contribute significantly to graphene properties.**

Our experiment was performed using a scanning tunneling microscope (STM) in UHV at $T$ = 4.8K to probe back-gated devices consisting of CVD-grown graphene[17] placed on top of boron nitride (BN) flakes[18] on a SiO$_2$/Si surface (see Supplementary Materials for methods). Utilization of BN substrates significantly reduces the charge inhomogeneity of graphene[19,20], thus allowing us to probe the intrinsic graphene electronic response to individual charged impurities. The charged impurities probed in this work were cobalt trimers constructed on graphene by atomically manipulating cobalt monomers with the tip of an STM[21] (cobalt atoms were deposited via e-beam evaporation onto low temperature graphene samples). Figs.1 a-f show the process of manipulating three cobalt monomers to create a single Co trimer on graphene (the detailed interior structure of the Co trimer cannot be resolved due to its instability under high current measurement).

Co trimers were used in this study because they form a robust, reproducible impurity whose charge can be toggled on and off through the use of a back-gate electrode. Co monomers (which can also be charge-toggled) were unsuitable for this study due to the fact that their charge state switches in the proximity of an STM tip[22] (thus leading to spatially inhomogeneous ionization features that mask the intrinsic graphene response to a Coulomb impurity of fixed charge). Co trimers are ideal because they can be prepared in different charge states (through back-gating) that are charge-stable in the proximity of an STM tip. This is shown in the spectroscopic data of Figs. 1g and 2. Fig. 1g shows STM spectra acquired with the STM tip held directly over a single Co trimer for two different back-gate voltages ($V_g$). Following the analysis of Ref. 22, the Co trimer impurity state marked **R** is seen to lie below the Fermi level ($E_F$) for $V_g = +14V$, in which case it is filled by an electron, while it lies above $E_F$ for $V_g = -2V$, in which case the electron has been removed from the trimer. The **R** state arises from the local cobalt-graphene hybridization and is undetectable at distances greater than $r_0 = 1.5$nm from the trimer center, demonstrating that short-range cobalt-graphene interactions end at $r_0$. The spectroscopic feature marked **S** arises due to tip-induced ionization of the trimer and thus confirms the charging nature of the impurity state **R**[22-25].

The gate-dependent charge states of the Co trimer were determined by performing dI/dV mapping of the surrounding graphene for different trimer charge configurations. Fig. 2a shows a map of the trimer in a bistable configuration with $V_g = +13V$. In this configuration the trimer exhibits a tip-induced ionization ring[22-25] (i.e., the trimer charge state changes by one electron depending on whether the tip is inside or outside the ring feature). In Fig. 2b the impurity state of the trimer (**R** in Fig. 1g) is well below $E_F$ ($V_g = +45V$), and is thus occupied by an electron. In this state the dI/dV map shows no significant structure outside the border of the trimer (i.e., for $r > r_0$, where $r$ is distance from the impurity center), implying that the trimer is in a stable charge-neutral state. In Fig. 2c, however, the impurity state **R** is well above $E_F$ ($V_g = -20V$), and so the trimer is charged here by an amount $Q = +1|e|$[22]. This charge state is consistent with the long-range radially symmetric contrast in dI/dV signal that is observed in the graphene beyond the border of the trimer. Similar behavior was reproducibly observed for more than 25 Co trimers (formed both naturally and through atomic manipulation) with 10 different tips calibrated against the Au(111) surface state.

Our ability to hold Co trimers in controlled, back-gate determined charge states allows us to measure the energy-dependent electronic LDOS of graphene around well-defined Coulomb impurities. Fig. 3a shows the results of such measurements taken at different distances from the center of a Co trimer in the charge state $Q = +1|e|$ ($V_g = -15V$). The dI/dV point spectra shown here have each been normalized by a different constant to account for the exponential change in conductivity that occurs at each measurement location as tip height is changed relative to the tip height at a distance far from the impurity[26,27] (see Supplementary Material for details). All the dI/dV spectra plotted in Fig.3a show a ~130meV wide gap-like feature at the Fermi level caused by phonon-assisted inelastic tunneling[28,29], and an additional minimum around $V_s = +0.13V$ which is associated with the Dirac point[29]. The primary difference between each spectrum is a systematic variation in the filled and empty state intensity as a function of distance from the trimer center. Empty state intensity above the Dirac point increases as the tip nears the trimer, while filled state intensity below the Dirac point correspondingly decreases.

Experimental dI/dV maps were obtained at different sample biases for the graphene surface surrounding a Co trimer in the $Q = +1|e|$ state. Figs.4a and c plot radial averaged dI/dV linescans measured as a function of distance from the Co trimer center. These curves have been normalized to account for the measured change in tip height at different spatial locations (see Supplementary Material). Fig.4a shows that the filled-state LDOS at energies below the Dirac point reduces near the Co trimer, but otherwise has very little spatial structure. The empty-state LDOS at energies above the Dirac point has very different behavior, as seen in Fig. 4c. Here the LDOS strongly increases as the trimer is neared, and a spatial oscillation that disperses with sample bias can be seen.

These experimental observations can be understood within a theoretical framework that incorporates the response of ultra-relativistic Dirac fermions to a Coulomb potential[6-9]. In this model the STM measures the LDOS of graphene quasiparticles as they arrange themselves around a charged impurity according to the physics of the massless Dirac equation. We calculated the expected LDOS of graphene at different distances away from a positively charged Coulomb impurity and compared this simulation to our distance-dependent, energy-resolved measurements of graphene LDOS around charged Co trimers. Calculations of the LDOS around a Coulomb impurity were carried out using the method of Ref. 8. This calculation assumes a 2D continuum Dirac model for undoped graphene in the presence of a Coulomb potential, and the only fitting parameter is the interband dielectric constant, $\varepsilon_g$, of undoped graphene. While the graphene used in our experiment has some finite doping ($\sim 5 \times 10^{11} \text{cm}^{-2}$), this model is still valid in proximity to the charged impurity since the effects of free-electron-like screening only become important for distances from the impurity greater than the screening length $\lambda \sim \frac{1}{k_F} = \frac{\hbar v_F}{\varepsilon_F} \sim 7\text{nm}^{30\text{-}32}$. For $r_0 < r \lesssim \lambda$ screening is dominated by the intrinsic interband contribution to graphene polarization ($\varepsilon_g$) whose main effect is to produce screening charge localized at the impurity center[30-32]. The overall charge of the Coulomb potential felt by graphene quasiparticles near the impurity within the screening length is thus reduced to $Q_{eff} = \frac{Q}{\varepsilon_g \varepsilon_s}$, where $Q$ is the bare impurity charge (in this case $Q = +1|e|$) and $\varepsilon_s = \frac{\varepsilon_{BN} + 1}{2} = 2.5$ is the average dielectric constant of the substrates surrounding the graphene ($\varepsilon_s$ arises from BN on one side and vacuum on the other).

We find that the data in Fig. 3a (which satisfies $r_0 < r \lesssim \lambda$) is best fit using $\varepsilon_g = 3.0 \pm 1.0$. Fig. 3b shows the corresponding theoretical dI/dV spectra calculated for the same distances from the charged impurity center as measured experimentally. These simulated spectra were obtained from the theoretical Dirac fermion LDOS (Fig.3b inset) by rigidly shifting the LDOS in energy to account for finite doping in the experiment, by broadening the LDOS with the finite quasiparticle lifetime, and by including phonon-assisted inelastic tunneling processes, all according to the method of Ref. 33 (estimated error in our extracted value of $\varepsilon_g = 3.0 \pm 1.0$ arises from the standard deviation in $\varepsilon_g$ values obtained from different sets of dI/dV spectra, see Supplementary Materials). Tip-induced band-bending was not included in our simulations because it is expected to be relatively small for the experimental doping levels used in this study (see Supplementary Materials for details).

As can be seen in Fig. 3b, the experimentally observed spatial dependence of the graphene LDOS around a charged impurity is well reproduced by the simulations. Namely, the LDOS of states above (below) the Dirac point are enhanced (reduced) as one moves closer to the impurity. This electron-hole asymmetry can be qualitatively understood as arising from the positive Coulomb potential of the Co trimer attracting negative charge carriers and repelling positive charge carriers. Our observation that the change in LDOS reduces as the energy nears the Dirac point is qualitatively different from the behavior expected for a conventional material having parabolic dispersion. This is because the electron-impurity interaction strength (the ratio of electronic potential energy over kinetic energy in the presence of a charged impurity) for graphene is independent of energy in contrast to the $\frac{1}{\sqrt{E}}$ behavior expected for this factor in conventional materials[31] (see supplement for a more detailed comparison between graphene and conventional materials). The lack of well-defined resonances in the graphene LDOS off the trimer indicates that no quasi-bound states are formed around the trimer, consistent with the theory of a subcritical graphene Coulomb impurity[6-8]. This is qualitatively different from traditional impurity systems involving massive fermions, in which case a Rydberg series of bound states always form around a Coulomb potential[2]. The value of $\varepsilon_g$ which is measured here by a direct "test-charge" method is close to the value calculated using the random phase approximation method[30,31] ($\varepsilon_{RPA} = 2.3$), but is significantly smaller than the value $\varepsilon_g \approx 15$ reported in Ref. 34. The small value of $\varepsilon_g$ implies that electron-electron interactions should play an important role in graphene[16], consistent with recent experiments reporting Fermi velocity renormalization[35,36].

Further confirmation of this theoretical interpretation of our data is seen in the simulated dI/dV linescan profiles of Figs. 4b and d for graphene surrounding a charged impurity. Here we observe a similar asymmetry in the electron-like and hole-like LDOS of graphene near a charged impurity. The theoretical linescans also reproduce another feature seen in the data, namely the appearance of dispersing spatial oscillations that are stronger in the electron-like LDOS above the Dirac point compared to the hole-like LDOS below it. These oscillations arise from the quantum interference of scattered Dirac fermions[37,38]. The electron-like interference amplitude is larger because electrons are pulled closer to the positive scattering center, and so experience a larger effective scattering phase shift than hole-like quasiparticles.

A difference between experiment and theory is that the theoretical electron-like LDOS has a long-range slope that is not seen in the data (i.e., for $r \gtrsim 6\text{nm}$). This can be explained by free-electron-like screening for $r > \lambda$ that arises from the finite doping of the experimental graphene.

In summary, we have employed tunable charge impurities to study the effect of a Coulomb potential on ultra-relativistic Dirac fermions over nanometer length scales. A strong position-dependent asymmetry is observed in the response of electron and hole quasiparticles to a charged impurity. These measurements directly demonstrate a regime where massless Dirac fermion physics departs markedly from typical massive fermion behavior, and can potentially explain differences in the mobility seen for electrons and holes in graphene transport measurements[10].

**Acknowledgements**: Research supported by the Office of Naval Research Multidisciplinary University Research Initiative (MURI) award no. N00014-09-1-1066 (graphene device preparation and characterization), by the Director, Office of Science, Office of Basic Energy Sciences of the US Department of Energy under contract no. DE-AC02-05CH11231 (STM instrumentation development and measurements), and by the National Science Foundation award no. DMR-0906539 (numerical simulations).

**Figure Captions**

**Fig. 1. (a to c)** STM topographs show the process of manipulating two Co monomers to form a Co dimer on a gated graphene device (atomic manipulation parameters: $V_s = +0.48V$, $I = 0.050nA$, $V_g = +45V$). **(d to e)** STM topographs show the process of manipulating a Co dimer to combine with a monomer for creation of a Co trimer on a gated graphene device (atomic manipulation parameters: $V_s = -0.46V$, $I = 0.060nA$, $V_g = -45V$). **(f)** Zoom-in STM topograph of a Co trimer on graphene (tunneling parameters: $V_s = +0.30V$, $I = 0.015nA$, $V_g = +30V$). **(g)** dI/dV spectra taken with tip directly above the center of a Co trimer on graphene for different back-gate voltages (initial tunneling parameters: $V_s = -0.15V$, $I = 0.020nA$).

**Fig. 2. (a)** dI/dV map of graphene near a Co trimer tuned to a bistable charge state using back-gate (tunneling parameters: $V_s = +0.30V$, $I = 0.018nA$, $V_g = +13V$). **(b)** dI/dV map of graphene near a Co trimer tuned to a stable $Q = 0$ charge state using back-gate ($V_s = +0.30V$, $I = 0.010nA$, $V_g = +45V$). **(c)** dI/dV map of graphene near a Co trimer tuned to a stable $Q = +1|e|$ charge state using back-gate ($V_s = +0.30V$, $I = 0.009nA$, $V_g = -20V$).

**Fig. 3. (a)** Normalized dI/dV spectra measured on graphene at different distances from Co trimer center when trimer is tuned to charge state $Q = +1|e|$ (initial tunneling parameters: $V_s = +0.30V$, $I = 0.015nA$, $V_g = -15V$, wiggle voltage $V_{rms} = 6mV$). Arrows indicate the direction towards the charged impurity. **(b)** Theoretically simulated normalized dI/dV intensity for graphene at same distances as in **(a)** for effective impurity charge

$$Q_{eff} = \frac{Q}{\varepsilon_g \varepsilon_s} = \frac{+1|e|}{3 \times 2.5} = +0.13|e|.$$ Inset: corresponding simulated bare LDOS of graphene calculated near impurity having $Q_{eff} = +0.13|e|$.

**Fig. 4. (a)** Experimental distance-dependent radial averaged normalized dI/dV linescans for graphene near a Co trimer with $Q = +1|e|$ charge (filled states, trimer center at $r = 0$). All curves are normalized by the value at $r = 13$nm and shifted vertically for easier viewing (experimental parameters: $V_g = -15$V and $(V_s, I) = (-0.50V, 55pA), (-0.40V, 55pA), (-0.35V, 40pA), (-0.30V, 40pA)$ and $(-0.25V, 30pA)$ from top to bottom under constant current feedback and with wiggle voltage $V_{rms} = 8$mV). **(b)** Theoretically simulated distance-dependent normalized dI/dV linescans for graphene near a charged impurity having $Q_{eff} = +0.13|e|$ (filled states, impurity center at $r = 0$). All theoretical curves are normalized by the value at $r = 13$nm. **(c)** Experimental dI/dV linescans for graphene near a Co trimer with $Q = +1|e|$ charge (empty states). Curves are plotted as in **(a)** (experimental parameters: $V_g = -15$V and $(V_s, I) = (0.48V, 58pA), (0.50V, 58pA), (0.55V, 55pA), (0.60V, 65pA), (0.70V, 90pA)$ and $(0.75V, 120pA)$ from top to bottom). **(d)** Theoretically simulated dI/dV linescans for graphene near a charged impurity having $Q_{eff} = +0.13|e|$ (empty states). Curves are plotted as in **(b)**.

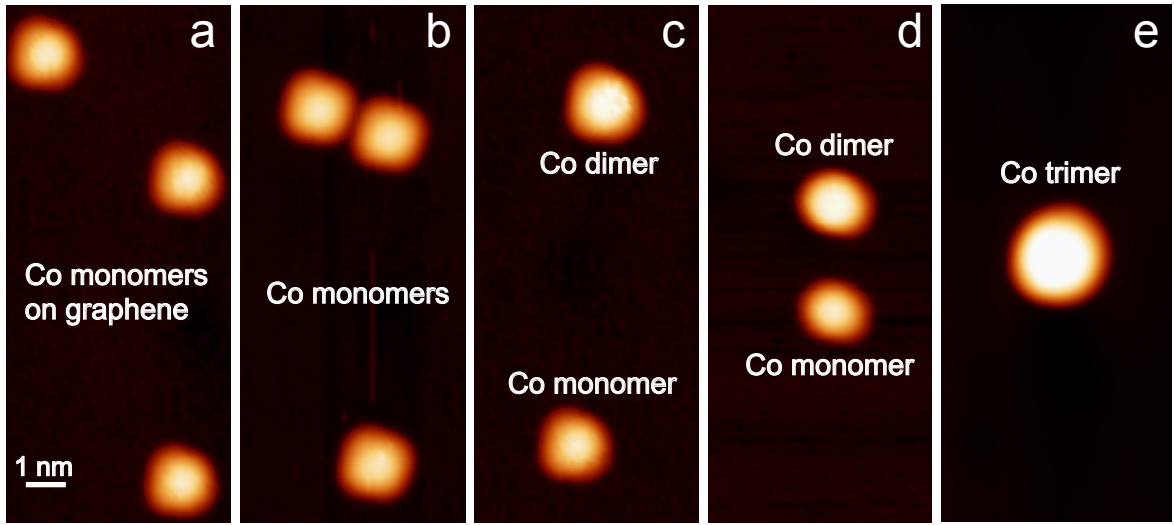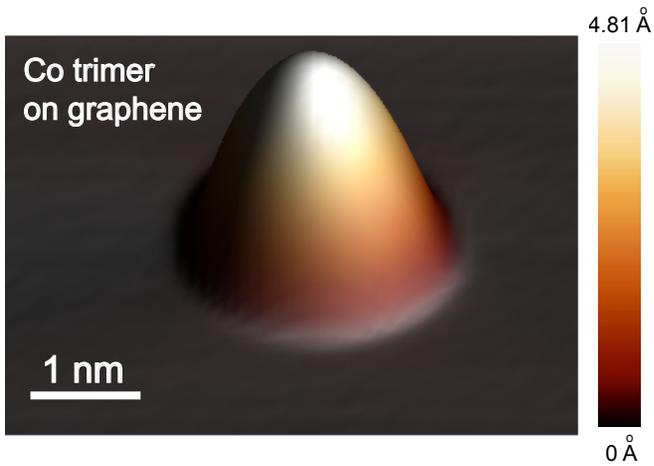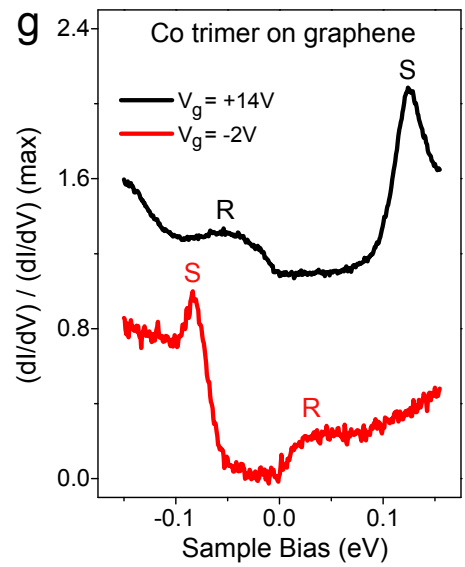

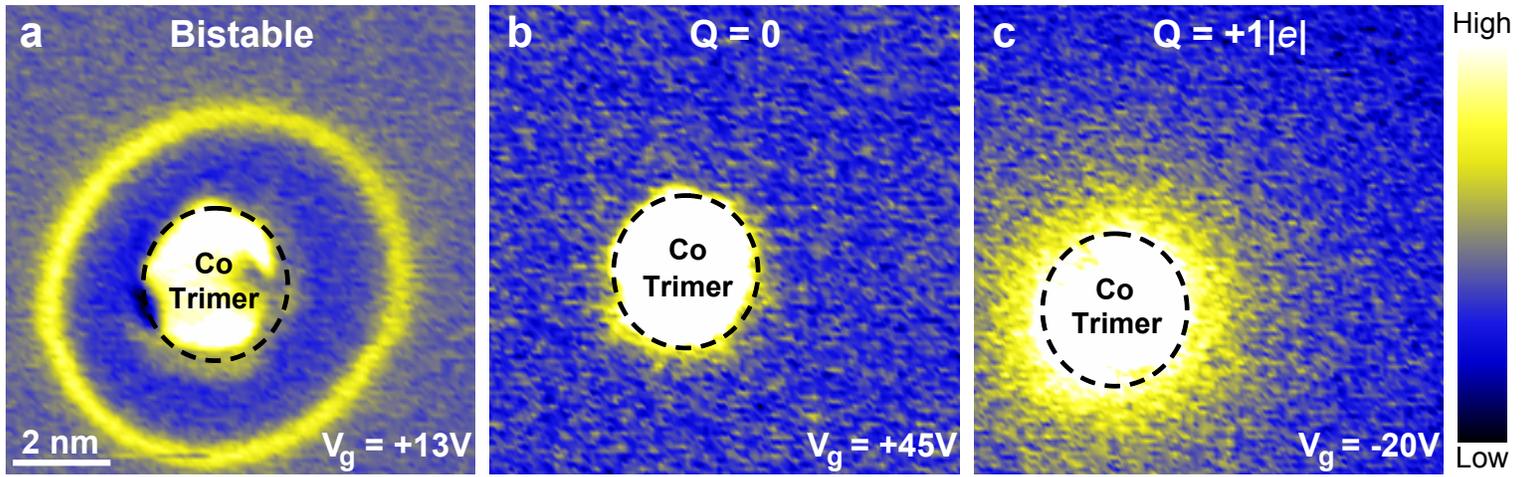

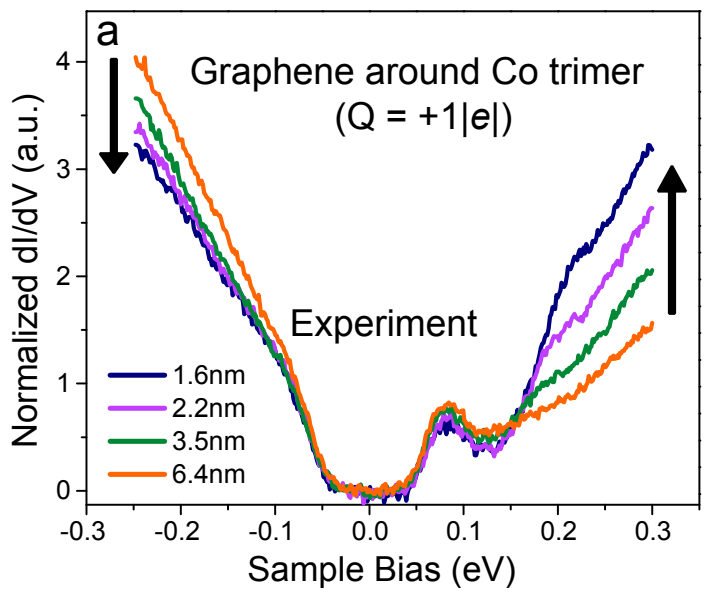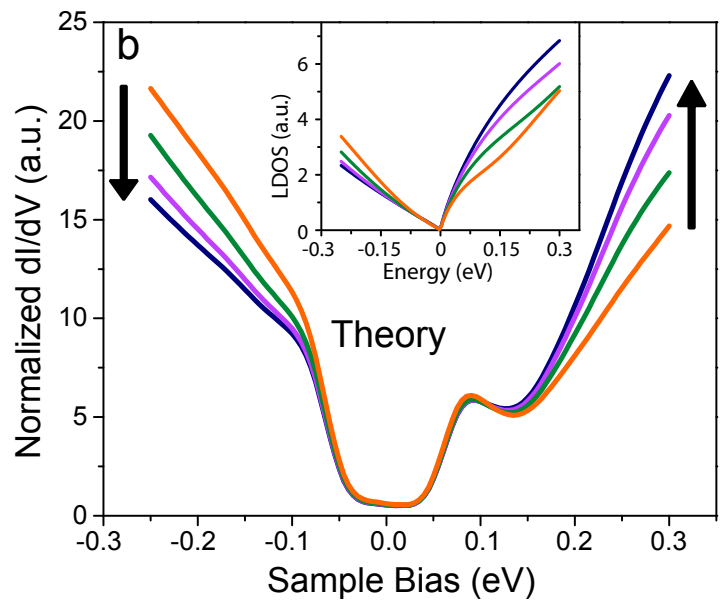

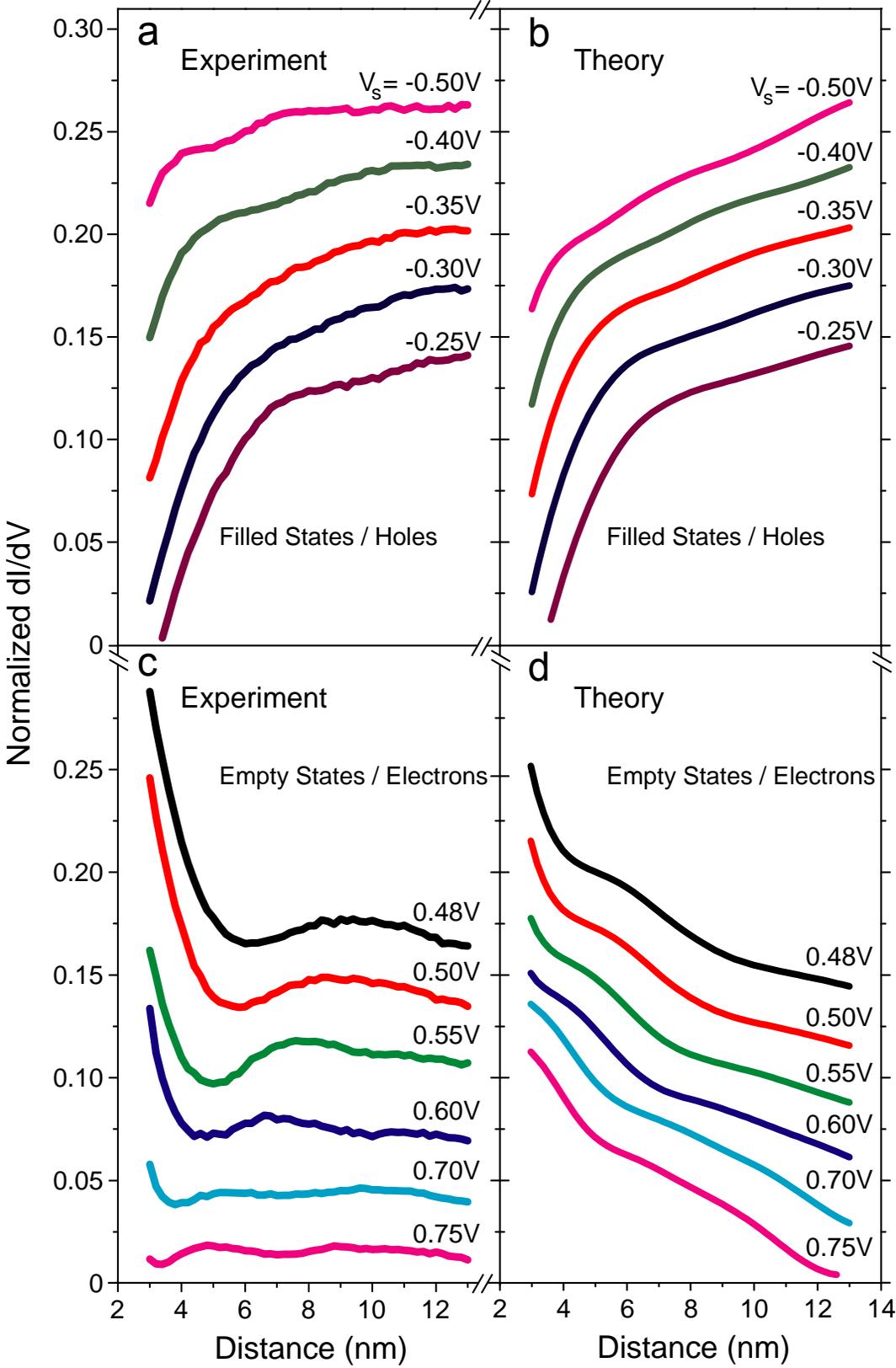

# Supplementary Information

**1. Experimental method**

The experiments were performed using an Omicron LT-STM under ultra-high vacuum (P < $10^{-10}$ Torr) at 4.8K. STM tips were calibrated against the Au(111) surface state before all measurements. Differential conductance (dI/dV) was measured by lock-in detection of the a.c. tunnel current modulated by a 6-9 mV (rms), 350-500 Hz signal added to the tunneling bias ($V_s$). The Co evaporator was first calibrated by evaporating Co atoms onto a Cu(111) surface and then checking the Kondo dip spectroscopic feature to indentify individual Co adatoms[1]. The graphene sample was grown by the CVD method described in Ref. 2. Boron nitride flakes (Momentive Company) were exfoliated onto heavily doped silicon wafers coated in 285nm thermal oxide. The graphene was placed on top of the BN / $SiO_2$ [3] and electrical contact was made by depositing Ti (10-nm thick)/Au (30-nm-thick) electrodes using a stencil mask technique. Samples were annealed in UHV at T ~ 400 °C for several hours to clean them before loading into the STM. Co atoms were then evaporated onto the graphene surface while holding the graphene at low temperature (<10K).

**2. Normalization method for dI/dV spectra**

The dI/dV spectra shown in Fig. 3a are normalized using Eq.1,

$$(\frac{dI}{dV})_{normalized} = (\frac{dI}{dV}) \times \exp(\frac{\Delta z}{\lambda}) \qquad (1)$$

where $\Delta z$ is the relative tip-height measured with respect to the tip-height far away from the impurity and obtained simultaneously with the spectra, and $\lambda = 0.45 \overset{0}{A}$ is the experimentally measured decay length measured for this tip material (PtIr)[4]. $\lambda$ thus takes into account the effects of graphene workfunction, tip workfunction, and barrier lowering effects. The important quantity here is the ratio between spectra measured at different locations and so any overall offset in the tip height will cancel out (note that Fig. 3 has arbitrary units). This normalization process removes the effect of the exponential fall-off in LDOS away from the surface[5] so that relative changes in normalized dI/dV spectra at different lateral distances from an impurity reflect the intrinsic in-plane response of the graphene surface[6,7]. This can be seen, for example, in Fig. S1 which shows both the bare and normalized dI/dV spectra at different lateral distances away from the Co trimer with two different initial tunneling set points. Different tunneling set points give rise to different distance-dependent dI/dV spectra due to the resulting different constant-current Z-height conditions (Figs. S1a and c). However, after applying the normalization process of Eq. 1, the normalized dI/dV spectra look similar for the different tunneling set points (Figs. S1b and d), and thus better reflect the intrinsic in-plane behavior of graphene.

This normalization process has also been applied to the radial averaged dI/dV spectra taken at different energies and as a function of distance for the Co trimer, as shown in Fig. 4 of the main text.

**3. Method for calculating LDOS of graphene near a charged impurity**

The theoretically simulated LDOS of graphene near a charged impurity shown in the inset of Fig. 3b is calculated using the same method as in Ref. 8. The simulation uses a 2D continuum model with linear dispersion for graphene, and the eigenstates for the Dirac equation under a Coulomb potential are then calculated (Eq. 4 in Ref. 8). The graphene electronic LDOS is then calculated using the eigenstate wavefunction expression (Eqs. 5, 10, 11 in Ref. 8). Additional effects of electron-phonon and electron-electron interactions on the STM dI/dV spectra are included using the method of Ref. 9 to obtain the simulated dI/dV spectra of Fig.3b.

**4. Fitting method used to extract dielectric constant of graphene ($\varepsilon_g$)**

The dielectric constant of graphene ($\varepsilon_g$) is extracted by comparing the experimental spectral intensity ratio between electron states ($V_b = 0.29 \pm 0.01$ volt) and hole states ($V_b = -0.24 \pm 0.01$ volt) at different lateral distances from the impurity to theoretical calculations using different values of $\varepsilon_g$ ($\varepsilon_g$ is the fitting parameter). To implement this in a statistically significant way, we define an asymmetry factor $\beta_{r_i}$ through Eq.2 (here $r_i$ refers to a specific distance from the impurity center):

$$\beta_{r_i} = (\frac{dI}{dV}\Big|_{V = +0.29 \pm 0.01 \text{ volt}})_{r_i} / (\frac{dI}{dV}\Big|_{V = -0.24 \pm 0.01 \text{ volt}})_{r_i} \qquad (2)$$

We then calculate the experimental ratio of asymmetry factors $\eta_{ij} = \beta_{r_i} / \beta_{r_j}$ for two different distances $r_i$ and $r_j$ from the impurity. The experimental value of $\eta_{ij}$ is then compared to the simulated $\eta_{ij}$ value (for the same two distances) calculated using $\varepsilon_g$ as a fitting parameter. $\varepsilon_g$ is then selected to optimize agreement between experimental and simulated $\eta_{ij}$. For a set of spectra taken at $n$ different distances from a Co trimer we perform this procedure for $n-1$ values of $\eta_{ij}$ where $r_j$ is fixed to be the furthest distance from the trimer and $r_i$ is varied over the other $n-1$ distances. All spectra used in this fitting procedure are within the range $r < \lambda$, where $\lambda$ is the screening length associated with free-electron-like screening (described in main text). This allows us to build up a histogram of values for $\varepsilon_g$.

This procedure was performed for 9 different sets of dI/dV spectra using different trimers and STM tips, and the resulting overall histogram of $\varepsilon_g$ is plotted in Fig. S2. The extracted value of $\varepsilon_g$ ($\varepsilon_g = 3.0 \pm 1.0$) is the average of this histogram, and the error bar in the extracted value is the standard deviation of the histogram. Possible origins of the scatter in fitted $\varepsilon_g$ values include random changes in the microstructure of the tip between different measurements and the proximity of randomly placed impurities and defects in the far field regime.

The asymmetry factor $\beta_{r_i}$ is used in this procedure because it is the fingerprint of the Coulomb potential strength for Dirac fermions. This fitting procedure helps eliminate electron-hole asymmetry in the spectra that might be induced by factors other than the charged impurity, such as the tip-density of states and energy-dependent tunneling barrier, since only the ratios between curves measured at different distances are fitted.

The experimental and theoretical asymmetry factors $\eta_{in} = \beta_{r_i} / \beta_{r_n}$ are plotted as a function of distance in Fig. S3. The theoretical curve is calculated using $\varepsilon_g = 3.0$. Two sets of experimental points are shown for different Co trimers and different tips.

**5. Intrinsic graphene dielectric constant obtained from random phase approximation calculations**

The intrinsic graphene dielectric constant ($\varepsilon_g$) can be calculated using the random phase approximation (RPA) method as in Ref. 10,11, resulting in the formula $\varepsilon_{RPA} = 1 + \frac{\pi \alpha}{2\varepsilon_s} = 2.3$. Here $\alpha = \frac{e^2}{\hbar v_F}$ is the fine structure constant of graphene and $\varepsilon_s = \frac{\varepsilon_{BN} + 1}{2} = 2.5$ is the average dielectric constant of the substrates surrounding the graphene.

**6. The effect of tip-induced band-bending**

The effect of tip-induced band-bending (TIBB) is estimated by comparing the position of features in the experimental and theoretical spatial derivative of the normalized dI/dV linescans, which are shown in Fig. S4. Figs. S4a and c show the spatial derivative of the experimental data shown in Figs. 4a and c respectively. Figs. S4b and d show the spatial derivative of simulated normalized dI/dV with biases picked to match the experiment (Figs. S4a and c).

The theoretical biases that best match the features between experiment and theory are slightly different from the actual biases used in the experiment because TIBB changes the effective doping of graphene at different applied biases. Therefore, the amount of TIBB can be estimated by the difference between the experimental (red dashed line) and effective sample biases (black dots), whose relation is plotted in Fig. S5. The extrapolated difference between the experimental and effective sample biases is very small for the bias range we used to extract the graphene dielectric constant ($V_b$ = -0.25V to +0.30V), and thus we conclude that TIBB is not significant enough to change any of the major conclusions of this study.

**7. Semi-classical analysis of Coulomb problem in graphene**

In this section we give a semi-classical derivation of the forbidden region for Dirac fermions in a Coulomb potential to explain why no graphene bound states or resonances can be formed in a weak Coulomb potential (i.e. the "subcritical" regime). We compare this with the solution of the non-relativistic hydrogen problem.

The total energy for a Dirac fermion near a Coulomb potential is $\varepsilon = v_F p - \frac{Ze^2}{r}$, where $v_F p$ is the kinetic energy of Dirac fermions in graphene and $-\frac{Ze^2}{r}$ is the Coulomb potential energy. After separating the radial and angular parts of momentum $p = \sqrt{p_r^2 + p_\theta^2}$, and substituting $p_\theta^2 = \frac{L^2}{r^2}$, one finds $p_r^2 = \frac{1}{v_F^2}(\varepsilon + \frac{Ze^2}{r})^2 - \frac{L^2}{r^2}$.

Following Ref. 12, the classically forbidden region corresponds to $p_r^2 < 0$. By quantizing $L = m\hbar$ where m is a half-integer[8] and examining the inequality, we find two different regimes depending on the magnitude of the charge Z. One is the subcritical regime and the other is the supercritical regime. In this work we focus on the subcritical regime.

The subcritical regime occurs when $Z < \frac{v_F |L|}{e^2}$. Here the forbidden region occurs for $r < \frac{|L|v_F - Ze^2}{|\varepsilon|}$ for states having $\varepsilon > 0$, and for $r < \frac{|L|v_F + Ze^2}{|\varepsilon|}$ for states having $\varepsilon < 0$. Therefore, in this regime, both electrons and holes will only scatter off a barrier at the center and no bound states or resonances can form.

The above behavior is qualitatively different from the non-relativistic hydrogen problem, because in the latter case bound states are formed for an arbitrarily weak Coulomb potential. The reason for this significant difference is that the linear dispersion in graphene causes the centrifugal barrier to have the form $L/r$ (where $L$ is the angular momentum), the same form as the Coulomb potential. Therefore (as shown above) when an attractive Coulomb potential is smaller than the centrifugal barrier, the total effective potential seen by an electron is always positive (i.e., barrier-like), and cannot host any bound states or resonances. For a normal parabolic dispersive material (such as a typical semiconductor) the centrifugal barrier has the form $L/r^2$ and decays faster than the $1/r$ Coulomb potential. In that case, no matter how weak the Coulomb potential is, it will always generate some attractive region and bound states will form there.

**8. Direct comparison between graphene and a 2D parabolic dispersive material in the presence of a charged impurity**

In order to further demonstrate that the behavior we observe near a charged impurity in graphene is unique for massless Dirac fermions, we have calculated the LDOS near a charged impurity in a 2D material with parabolic dispersion and we compare that to the calculated LDOS near a similarly charged impurity in graphene (in the subcritical regime).

We consider a 2D material in which the conduction band and valence band are separated by a 1eV bandgap. Both bands have parabolic dispersion with an effective mass of $0.05m_e$ ($m_e$ is the electron mass), and the dielectric constant, $\varepsilon$, is chosen to be 2.5 (the behavior described here does not depend significantly on these parameters). The LDOS of the conventional 2D material near a positively charged impurity is calculated by solving the eigenfunctions of the Schrödinger equation:

$$(-\frac{\hbar^2}{2m}\nabla^2 - \gamma\frac{Ze^2}{\varepsilon r})\psi(r,\theta) = E\psi(r,\theta) \quad \text{where} \quad \gamma = +1 \quad \text{for electrons in the conduction}$$

band and $\gamma = -1$ for holes in the valence band.

The final results are plotted in Fig. S6a along with a comparison to the corresponding simulation for graphene (Fig. S6b). The energy-dependent LDOS for both cases is plotted for different distances from the positive impurity. For both the semiconductor and graphene, the LDOS without the impurity is plotted as a dashed reference line. The most obvious difference between graphene and the semiconductor is the appearance of bound states in the bandgap of the 2D parabolic material in analogy to the Rydberg series for hydrogen (no bound states or resonances are seen for graphene). The 2D semiconductor bound state energy levels are $E_n = -\frac{Z^2 m e^4}{2n^2 \varepsilon^2 \hbar^2}$ measured from the bottom of the conduction band, where $n = \frac{1}{2}, \frac{3}{2}, \frac{5}{2}, \cdots$ (these energy levels are different than standard hydrogen because the electrons are confined to a plane).

We now compare the LDOS of the continuum states for the two cases. There are two major qualitative differences for the continuum states near the charged impurity between graphene and the semiconductor, as seen in Fig. S6. The first major difference is that the LDOS of graphene displays a simpler electron-hole asymmetry near the positively charged impurity, which is illustrated by the fact that the LDOS at energies above (below) the Dirac point is always bigger (smaller) than the bare graphene LDOS at all positions. The semiconductor displays a very different behavior. While a general electron-hole asymmetry is seen near the positively charged impurity for the semiconductor, the behavior of this asymmetry is very different from what we see in graphene. For example, at the bottom of the conduction band there are strongly fluctuating variations in LDOS, and the LDOS of the valence band becomes bigger than the bare LDOS at some energies and distances (something that does not happen for subcritical charged impurity in graphene). The second major difference between graphene and the semiconductor is that the LDOS of graphene at the Dirac point is unaffected by the charged impurity, and impurity-induced changes in graphene LDOS only become larger as one moves further in energy from the Dirac point. This is in marked contrast to the LDOS variations of the semiconductor near a charged impurity, which are strongest near the band edges and which become weaker as one moves further in energy from the band edges.

The physical origin of the differences in continuum state behavior can be traced to the unique electronic structure of graphene. The reason for the first difference is that for charge carriers in the semiconductor, the impurity effective potential has an attractive region and so incoming and outgoing waves experience more interference near the impurity, resulting in stronger LDOS oscillations. This does not happen in subcritical graphene since the effective potential seen by Dirac fermions is always repulsive. The reason for the second difference is that the electron-impurity interaction strength $r_s$, i.e. the ratio of electronic potential energy over kinetic energy in the presence of a charged impurity, scales differently with energy for a semiconductor versus graphene. For an electron with wavenumber k, the potential energy can be estimated by taking the de Broglie wavelength 1/k as a typical distance from the impurity, which gives the potential energy $U \sim \frac{e^2}{r} \sim e^2 k$. For a semiconductor, the kinetic energy behaves as $K \sim \frac{\hbar^2 k^2}{m}$, and so the interaction strength behaves as $r_s = \frac{U}{K} \sim \frac{e^2 m}{\hbar^2 k} \sim \frac{1}{\sqrt{E}}$. Therefore, when combined with a constant density of states the impurity has biggest impact at low energies for the semiconductor (consistent with the simulation). For graphene, on the other hand, the kinetic energy behaves as $K \sim \hbar v_F k$, and so the interaction strength behaves as $r_s = \frac{U}{K} \sim \frac{e^2}{\hbar v_F}$, and is energy independent. Therefore, when combined with a linear density of states the impurity has biggest impact at high energies in graphene (completely opposite to the behavior of the semiconductor).

**Figure Captions**

**Fig. S1. (a)** Bare dI/dV spectra measured on graphene at different distances from Co trimer center when trimer is tuned to charge state Q=+1|e| (initial tunneling parameters: $V_s$ = +0.3V, I = 0.015nA, $V_g$ = -15V, wiggle voltage $V_{rms}$ = 6mV). **(b)** Normalized dI/dV spectra obtained by normalizing the bare dI/dV spectra in **(a)** according to the text. **(c)** Bare dI/dV spectra measured on graphene at different distances from Co trimer center when trimer is tuned to charge state Q=+1|e| (initial tunneling parameters: $V_s$ = -0.25V, I = 0.015nA, $V_g$ = -15V, wiggle voltage $V_{rms}$ = 6mV). **(d)** Normalized dI/dV spectra obtained by normalizing the bare dI/dV spectra in **(c)** according to the text.

**Fig. S2** Histogram of the fitted value of intrinsic graphene dielectric constant using the method described in the text.

**Fig. S3** Comparison between the experimental and theoretical distance dependence of the asymmetry factor. The theoretical curve is calculated using $\varepsilon_g = 3.0$. The blue circles are obtained from the set of spectra shown in Fig. 3a of the main text. The black squares are obtained from a different set of spectra measured on a different Co trimer and with a different tip.

**Fig. S4 (a and c)** Experimental spatial derivative of the normalized dI/dV linescans near a Co trimer with $Q = +1|e|$ charge. **(b and d)** Theoretically simulated spatial derivative of normalized dI/dV linescans for graphene near a charged impurity having

$Q_{eff} = +0.13|e|$, with biases chosen to fit the data in **(a)** and **(c)**.

**Fig. S5** The effective sample biases fitted by comparing the experiment and theory in Fig. S4 as a function of experiment sample biases, which are plotted as black dots. The red dashed line plots the line when effective and experimental sample biases are the same (i.e., when there is no TIBB).

**Fig. S6 (a)** Simulation of LDOS at different distances from a charged impurity in a model 2D parabolic dispersive material (the black dashed line represents the bare LDOS without the impurity). **(b)** Simulation of LDOS at the same distances as (a) from a similarly charged impurity in graphene (the black dashed line represents the bare graphene LDOS without the impurity).

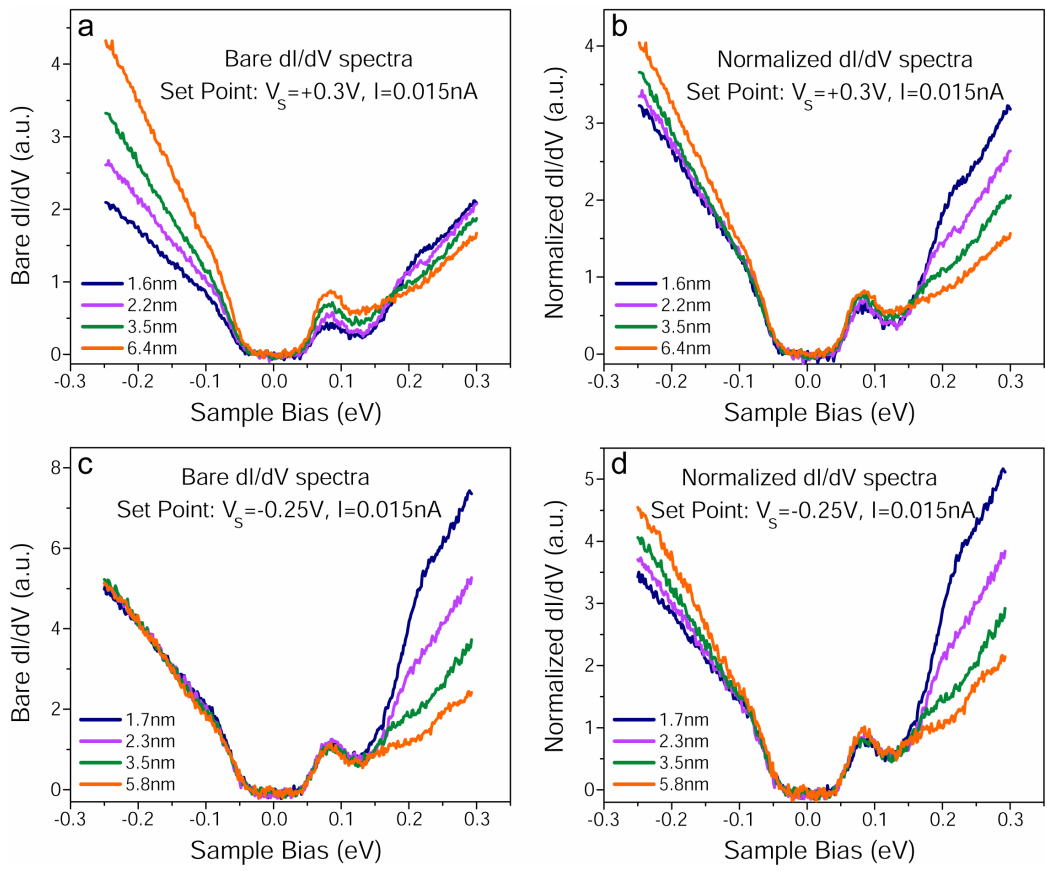

Fig. S1

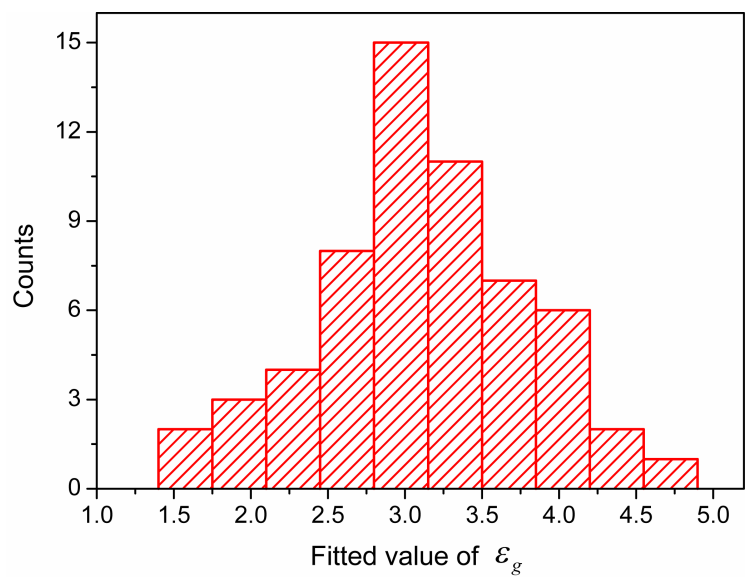

Fig. S2

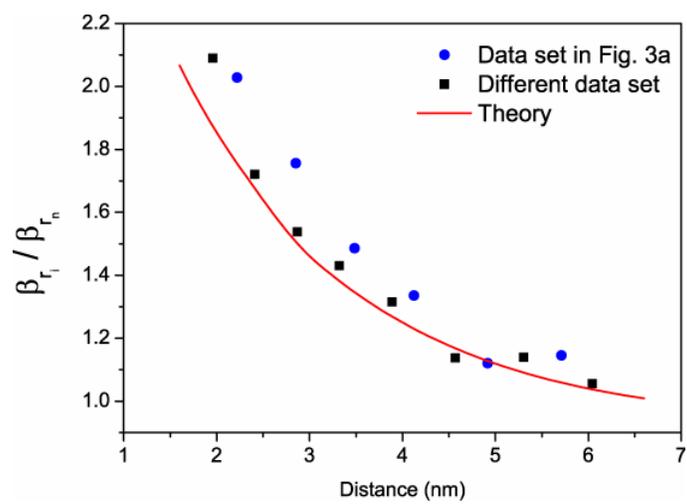

Fig. S3

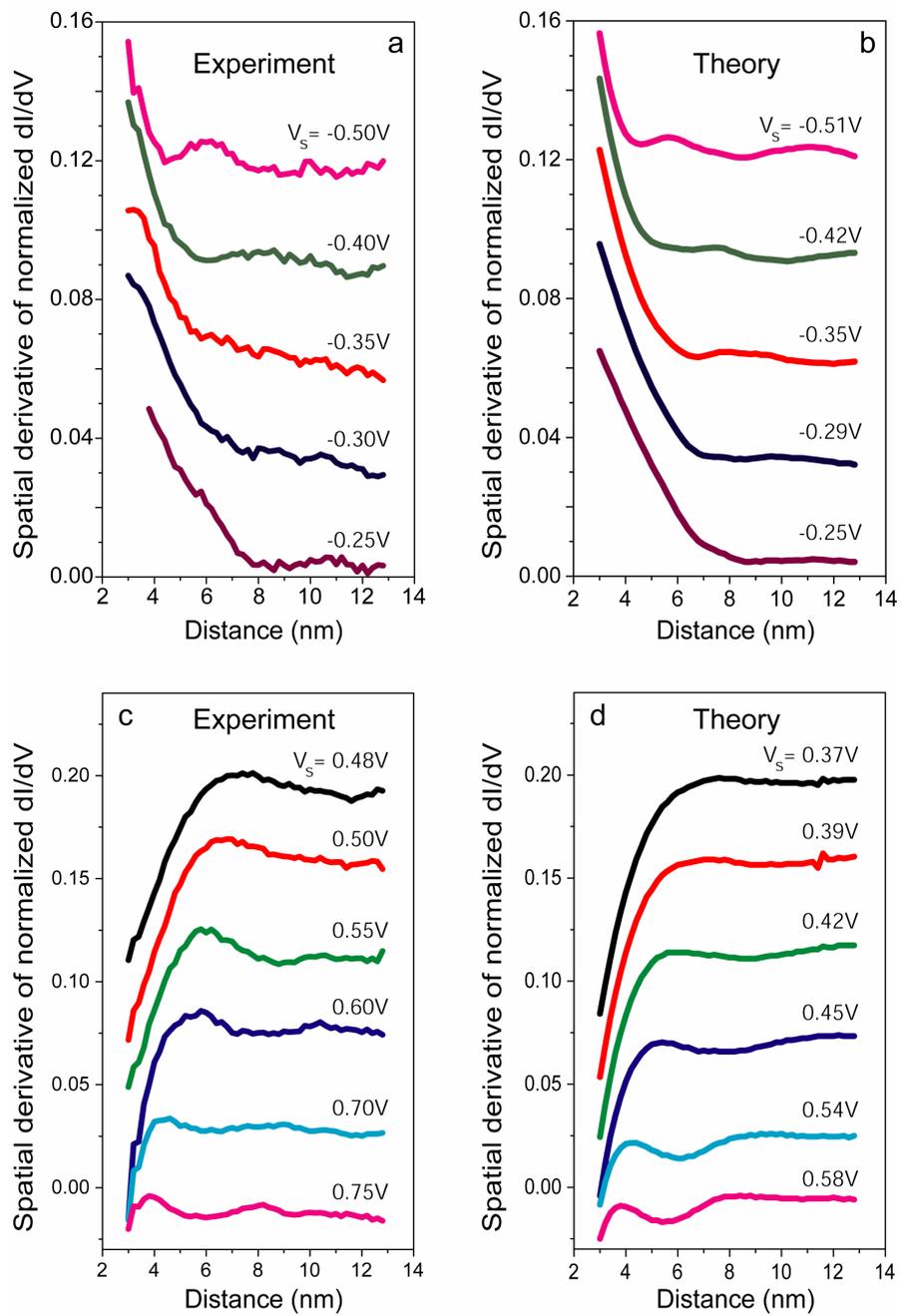

Fig. S4

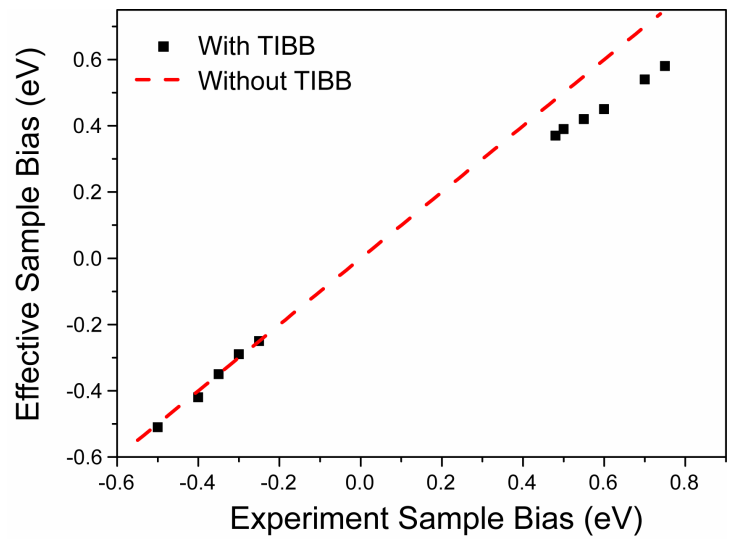

Fig. S5

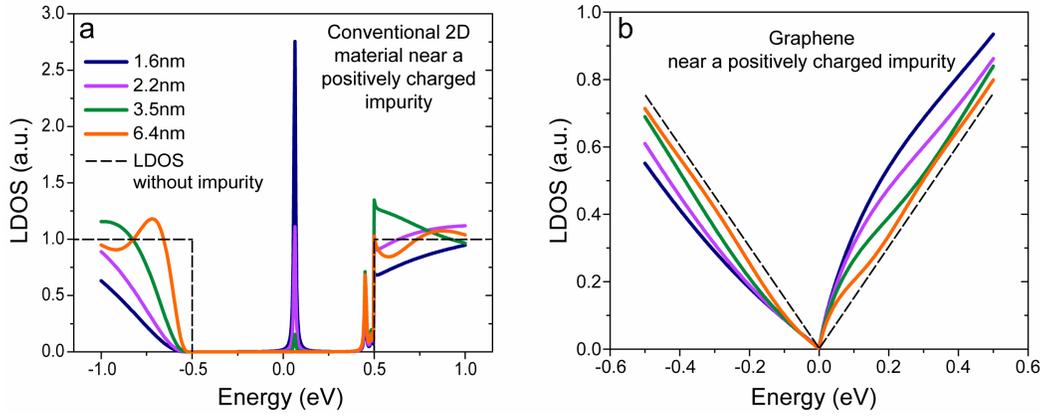

Fig. S6